%% file: main.tex
\documentclass[sigconf]{acmart}

\usepackage{threeparttable}
\usepackage{makecell}
\usepackage{multirow}
\usepackage[most]{tcolorbox}

\usepackage{listings}
\usepackage{xcolor}

\usepackage{algorithm}
\usepackage{algpseudocode}
\usepackage{float}

\algnewcommand\algorithmicinput{\textbf{Input:}}
\algnewcommand\Input{\item[\algorithmicinput]}

\algnewcommand\algorithmicoutput{\textbf{Output:}}
\algnewcommand\Output{\item[\algorithmicoutput]}

\algnewcommand\algorithmicattributes{\textbf{Special Attributes:}}
\algnewcommand\Attributes{\item[\algorithmicattributes]}

\definecolor{eclipseStrings}{RGB}{42,0.0,255}
\definecolor{eclipseKeywords}{RGB}{127,0,85}
\colorlet{numb}{magenta!60!black}

\lstdefinelanguage{json}{
    basicstyle=\normalfont\ttfamily,
    commentstyle=\color{eclipseStrings}, 
    stringstyle=\color{eclipseKeywords}, 
    numbers=left,
    numberstyle=\scriptsize,
    stepnumber=1,
    numbersep=8pt,
    showstringspaces=false,
    breaklines=true,
    frame=lines,
    backgroundcolor=\color{gray}, 
    string=[s]{"}{"},
    comment=[l]{:\ "},
    morecomment=[l]{:"},
    literate=
        *{0}{{{\color{numb}0}}}{1}
         {1}{{{\color{numb}1}}}{1}
         {2}{{{\color{numb}2}}}{1}
         {3}{{{\color{numb}3}}}{1}
         {4}{{{\color{numb}4}}}{1}
         {5}{{{\color{numb}5}}}{1}
         {6}{{{\color{numb}6}}}{1}
         {7}{{{\color{numb}7}}}{1}
         {8}{{{\color{numb}8}}}{1}
         {9}{{{\color{numb}9}}}{1}
}

\definecolor{codegreen}{rgb}{0,0.6,0}
\definecolor{codegray}{rgb}{0.5,0.5,0.5}
\definecolor{codepurple}{rgb}{0.58,0,0.82}
\definecolor{backcolour}{rgb}{0.95,0.95,0.92}

\lstdefinestyle{myStyle}{
    belowcaptionskip=1\baselineskip,
    breaklines=true,
    frame=none,
    numbers=none, 
    basicstyle=\footnotesize\ttfamily,
    keywordstyle=\bfseries\color{green!40!black},
    commentstyle=\itshape\color{purple!40!black},
    identifierstyle=\color{blue},
    backgroundcolor=\color{gray!10!white},
}

\AtBeginDocument{%
  }

\setcopyright{acmlicensed}
\copyrightyear{2025}
\acmYear{2025}
\acmDOI{XXXXXXX.XXXXXXX}

\acmConference[Conference acronym 'XX]{Make sure to enter the correct
  conference title from your rights confirmation email}{June 03--05,
  2018}{Woodstock, NY}
\acmISBN{978-1-4503-XXXX-X/18/06}

\newcommand\tool[1]{\emph{PTMPicker}}

\newcommand\etal[1]{\textit{et al.}}
\newcommand\ie[1]{\textit{i.e.}}
\newcommand\eg[1]{\textit{e.g.}}
\newcommand\perse[1]{\textit{per se}}

\begin{document}

\title{\tool{}: Facilitating Efficient Pretrained Model Selection for Application Developers}



\author{Pei Liu}
\email{peiliu@cqu.edu.cn}
\affiliation{%
  \institution{Chongqing University}
  \country{China}
}

\author{Terry Zhuo}
\email{terry.zhuo@monash.edu}
\affiliation{%
  \institution{Monash University and CSIRO's Data61}
  \country{Australia}
}

\author{Jiawei Deng}
\email{dengjiaweibryan@gmail.com}
\affiliation{%
  \institution{CSIRO's Data61}
  \country{Australia}
}

\author{Zhenchang Xing}
\affiliation{%
  \institution{CSIRO's Data61}
  \country{Australia}}
\email{zhenchang.xing@data61.csiro.au}

\author{Qinghua Lu}
\affiliation{%
  \institution{CSIRO's Data61}
  \country{Australia}}
\email{qinghua.lu@data61.csiro.au}

\author{Xiaoning Du}
\affiliation{%
  \institution{Monash University}
  \country{Australia}}
\email{Xiaoning.Du@moansh.edu}

\author{Hongyu Zhang}
\email{hyzhang@cqu.edu.cn}
\affiliation{%
  \institution{Chongqing University}
  \country{China}
}


\begin{abstract}
The rapid emergence of pretrained models (PTMs) has attracted significant attention from both Deep Learning (DL) researchers and downstream application developers. 
However, selecting appropriate PTMs remains challenging because existing methods typically rely on keyword-based searches in which the keywords are often derived directly from function descriptions. 
This often fails to fully capture user intent and makes it difficult to identify suitable models when developers also consider factors such as bias mitigation, hardware requirements, or license compliance. 
To address the limitations of keyword-based model search, we propose~\tool{} to accurately identify suitable PTMs. 
We first define a structured template composed of common and essential attributes for PTMs and then~\tool{} represents both candidate models and user-intended features (\ie{}, model search requests) in this unified format. 
To determine whether candidate models satisfy user requirements, it computes embedding similarities for function-related attributes and uses well-crafted prompts to evaluate special constraints such as license compliance and hardware requirements. 
We scraped a total of 543,949 pretrained models from Hugging Face to prepare valid candidates for selection. 
\tool{} then represented them in the predefined structured format by extracting their associated descriptions. 
Guided by the extracted metadata, we synthesized a total of 15,207 model search requests with carefully designed prompts, as no such search requests are readily available. 
Experiments on the curated PTM dataset and the synthesized model search requests show that~\tool{} can help users effectively identify models, with 85\% of the sampled requests successfully locating appropriate PTMs within the top-10 ranked candidates. 
\end{abstract}

\ccsdesc[500]{Software and its engineering~Application specific development environments}
\ccsdesc[500]{Applied computing~Document metadata}

\keywords{Pretrained Models, Model Selection, Metadata, Template, Hugging Face}


\maketitle

\input{Sections/intro}
\input{Sections/motivation}

\input{Sections/approach}

\input{Sections/evaluation}
\input{Sections/discussion}

\vspace{-4pt}
\section{Conclusion}
In this paper, we propose~\tool{} to identify suitable pretrained models for downstream application developers, including those who may not have extensive AI expertise. 
\tool{} first converts the PTM associated descriptions into structured metadata representation via 
GPT-4, while model users also express their intended features of target models in the same format. 
Finally, it computes the embedding similarity for the same functional-related metadata fields between candidate PTMs and model users' request to find matches. 
With regard to the special 
requirements in the search requests, such as limitations and biases, specific prompts are designed to determine whether the candidates satisfy or not. 
In addition, we curated a candidate PTM dataset from the most popular model hub, Hugging Face, containing a total of 543,949 pre-trained models, and synthesized 15,207 model search requests as we still do not have such a dataset for our research, which can not only fulfill our research intention but also provide a good research basis for fellow researchers. 
Experiments on randomly selected samples show that~\tool{} achieves an accuracy of 85\% in identifying qualified PTMs within the top-10 ranked candidates.

\textbf{Data Availability:}
The source code and datasets are all made publicly available in our artifact package via the following link:
\url{https://zenodo.org/records/15958951}


\bibliographystyle{ACM-Reference-Format}
\bibliography{main}
\end{document}

%% file: Sections/intro.tex
\section{Introduction}

PreTrained Models (PTMs) refer to neural network models that have been previously trained on large-scale datasets and saved for future reuse. 
In recent years, the existing literature on PTMs discusses the core issues and risks of the PTM supply chain and hubs~\cite{mitchell2019model,jiang2022empirical,castano2023exploring,kathikar2023assessing} and the detailed utilization of specific PTMs for special tasks~\cite{schmidhuber2015deep,antoniou2017data,dube2019automatic,xia2022systematic,qiu2020pre}, such as molecular property prediction in chemistry. 
In addition to the literature on PTM supply chain risks and domain-specific applications, there is also a stream of research~\cite{bolya2021scalable,you2021logme,zhang2023model,liu2023task,davis2023reusing} dedicated to the identification and selection of suitable models. 
Specifically, Zhang~\etal{}~\cite{zhang2023model} proposed Model Spider, which selects suitable pretrained models for specific tasks by computing and ranking the similarity between the candidate model token representation and the downstream task token representation. 
Model Spider better serves researchers and developers who are interested in the nuanced differences among various PTMs and seek to explore their performance across specific evaluation metrics. 
Liu~\etal{}~\cite{liu2023task} introduces MLTaskKG to recommend ML/DL libraries/models for application developers based on the specifications of AI tasks specified by them. 
MLTaskKG primarily focuses on AI task descriptions while overlooking the limitations of recommended models or libraries, such as potential biases, safety concerns, and license compatibility. 

The surge in PTMs has led to the emergence of Hugging Face~\cite{hfsite:2025,hfwiki:2025} as the~\textit{de facto} standard model hosting platform. 
Hugging Face, an online Git-based platform, hosted more than 543,000 diverse pretrained models at the time of our study, and the number of models continues to grow rapidly. 
This centralized repository could obviate the need for researchers to manually retrieve models from other heterogeneous hosting platforms. 
Although the platform provides model specifications, developers\footnote{For simplicity, unless otherwise specified, the term developers hereafter refers to downstream application developers in this study.} still face challenges in finding models that align with their requirements. 
In practice, application developers are often concerned not only with the intended functionality of potential target PTMs but also with aspects such as license compliance, model size, and interface usability. 
When searching using functional keywords or high-level task categories, an overwhelming number of candidate models are typically returned. 
Developers then need to devote considerable attention to manually and carefully examining and iterating all the candidate models, which is both labor-intensive and time-consuming. 

Therefore, it is essential to develop an effective and automated model selection approach that can rapidly and accurately identify models fulfilling task requirements while also satisfying specific constraints, such as license compatibility and safety considerations.  
This, in turn, can significantly accelerate downstream application development by reducing manual effort and improving selection precision. 
As a further benefit, it may also incentivize model developers to offer clearer and more informative specifications for their models, thereby improving the models' discoverability and adoption during selection. 

To this end, we propose a prototype,~\tool{}, to assist application developers in selecting suitable pretrained models based on our predefined template. 
The template is derived from existing literature and specifications provided by model developers when publishing, and is designed to comprehensively capture essential properties such as functionality, limitations, biases, and licensing. 
Specifically, the specifications of each model are transformed by~\tool{} into a structured representation defined by the predefined template, hereafter referred to as a model-populated instance. 
Downstream developers can likewise articulate their requirements for the pretrained model by populating the corresponding attribute fields in this template, hereafter referred to as a developer-populated instance or search request. 
To select appropriate models,~\tool{} first computes embedding similarities between the function-related fields specified in the developer-populated instance and the corresponding fields in the model-populated instances. 
It then ranks the pretrained models based on the calculated similarity values. 
In addition, for each special requirement, such as specific license specifications and hardware constraints,~\tool{} employs a well-crafted proprietary prompt to assess whether each high-ranking PTM satisfies the particular constraint. 
Finally, the pretrained models that satisfy all these special requirements and rank in the top 10 are returned to developers for further selection, thereby eliminating the need to spend considerable time manually inspecting a large number of potential models. 

To evaluate the effectiveness of~\tool{}, a large number of pretrained models along with their associated descriptions, and developer-specified model search requests are necessary. 
However, such pretrained models and developer-specified requests are not readily available in the research community. 
We therefore first constructed the dedicated dataset of pretrained models by scraping Hugging Face, and created a substantial set of high-quality model search requests by adopting the concept of mutation from software testing. 
Specifically, we automatically scraped pretrained models and their descriptions, and then converted these descriptions into well-structured metadata representations by extracting and populating the predefined template attributes. 
We subsequently mutated the resulting model-populated instances using deliberately crafted prompts---for example, by altering the functional requirement from text-to-image to image-to-text---in order to expand the dataset of model search template instances (\ie{}, model search requests). 
Experiments on model search requests and the curated large-scale PTM dataset show that~\tool{} could effectively help model users find appropriate models, with 85\% of the 100 sampled requests successfully identifying suitable PTMs within the top-10 ranked candidates. 


In general, the main contributions of this research are as follows.
\begin{itemize}
    \item \textbf{\tool{}, PTM Selection Approach:} We proposed an effective approach to assist developers automatically select suitable pretrained models from the vast pool of available models. 
    \item \textbf{PTM Dataset:} We curated a PTM dataset comprising 543,949 publicly available pretrained models sourced from Hugging Face, the most widely used model hub. 
    \item \textbf{PTM Search Request Dataset:} We introduced a systematic approach to automatically synthesize model search requests that reflect the needs of downstream application developers, and we also constructed a corresponding dataset comprising 15,207 such requests. 
\end{itemize}

%% file: Sections/motivation.tex
\section{Background \& Motivation}


Compared with training models from scratch, pretrained models (PTMs) substantially reduce cost and time requirements, and, when coupled with post-training compression, such as quantization—enable deployment on resource-constrained devices \cite{egashira2024exploiting}. 
This efficiency lowers barriers for developers by allowing goal-oriented reuse through inference with minimal integration effort, making PTM-centric workflows attractive across domains from general AI systems to embedded/IoT software~\cite{MSurvey9,MSurvey5}. 
At the same time, widespread PTM adoption intersects with security considerations in which LLM-based components introduce new attack surfaces and must be assessed against evasion and supply-chain risks documented in recent surveys and empirical studies \cite{MSurvey6,MSurvey7,MSurvey8,MTech1}.

The emergence and widespread adoption of PTMs has attracted significant research attention, with some researchers focusing on the development and analysis of model documentation, commonly referred to as model cards.  
Mitchell~\etal{}~\cite{mitchell2019model} first proposed model cards for two types of Deep Learning (DL) models: one trained to identify smiling faces in images and the other trained to detect offensive comments in natural language texts. 
These model cards were intended to promote model transparency among not only users but also developers and stakeholders. 
Going one step further, Crisan~\etal{}~\cite{crisan2022interactive} investigated interactive model cards, which extend traditional static cards with features for richer documentation exploration and direct model interaction. 
Drawing on insights from interviews with DL/ML experts and non-experts, their analysis highlights the importance of thoughtful design and interactivity, offering guidelines for more human-centered model cards. 

Meanwhile, some other investigators have focused on the development and distribution of the PTMs themselves. 
Han~\etal{}~\cite{han2021pre} comprehensively reviewed the past and present developments in PTMs. 
They analyzed recent advances driven by the availability of large training datasets and rapidly increasing computational power, and proposed several promising directions for future research. 
Jiang~\etal{}~\cite{jiang2022empirical} first conducted a comprehensive empirical study in the pretrained model supply chain and 
analyzed eight popular model hubs that host PTMs and facilitate sharing between model developers and users. 
Their study revealed that the PTM supply chain plays a vital role in the DL supply chain and the existing security approaches for the PTM supply chain are insufficient. 

Furthermore, Zhang~\etal{}~\cite{zhang2023model} introduced Model Spider to help users select appropriate pretrained models by computing and ranking similarity between token representations of potential models and the user's intended features. 
Li~\etal{}~\cite{li2023guided} also proposed a similar approach for recommending appropriate PTMs to users, which we refer to as ``Guided Recommendation'' for brevity. 
They addressed the model selection challenge by formulating it as a model recommendation problem. 
Specifically, they leveraged related information, such as the features and labels of the models’ training datasets, along with their objective tasks, to compute guidance scores that help recommend suitable PTMs to users. 
Both approaches focus on selecting appropriate pretrained models for downstream tasks. 
However, Guided Recommendation emphasizes incorporating task-specific guidance to improve PTM recommendation, whereas Model Spider focuses on learning matching patterns between PTMs and tasks to enable effective selection. 
Nevertheless, the effectiveness of both Model Spider and Guided Recommendation depends on the availability of the pretrained models' training datasets, the datasets for the downstream tasks and the downstream task objectives. 
In downstream application development scenarios, such datasets and precise task objectives are not always accessible to downstream application developers. 
As a result, these approaches are particularly well-suited for researchers and developers who have expertise in AI and are interested in exploring subtle differences among pretrained models, selecting models with the best performance across standardized evaluation criteria, and further fine-tuning them for specific application contexts. 

Moreover, Liu~\etal{}~\cite{liu2023task} proposed MLTaskKG, a system designed to recommend machine learning (ML) and deep learning (DL) libraries or models to application developers based on the AI task specifications they provide. 
MLTaskKG constructs a comprehensive knowledge graph by aggregating publicly available pretrained models and libraries from diverse sources such as PapersWithCode~\cite{paperswithcode:2025} and published literature, and modeling relationships between AI tasks, ML/DL models, and their underlying implementations. 
This knowledge graph is subsequently leveraged to recommend appropriate ML/DL libraries by matching developer-specified tasks, model characteristics, and implementation requirements. 
The approach is particularly advantageous for downstream application developers who have limited ML/DL expertise. 
However, MLTaskKG mainly emphasizes functional-oriented characteristics while paying limited attention to potential limitations of the recommended models or libraries, such as issues related to bias, safety, or license compatibility. 
Moreover, the project is no longer actively maintained and we are unable to replicate the system due to the lack of up-to-date implementation. 


In practice, obtaining appropriate PTMs from Hugging Face for downstream application developers with limited AI expertise involves at least two consecutive steps. 
The first step is to select one task from the total of 44 predefined model task tags, such as Visual Question Answering and Image Classification. 
By selecting a predefined task tag, developers are presented a large number of available PTMs. 
Additionally, they can also sort the filtered PTMs by trending status, number of downloads, number of likes, recency of updates or creation on the platform. 
After selecting a task and sorting the resulting PTMs, developers are then required to read through filtered model cards, even further their related published papers, carefully, one by one, in order to identify the best fits, which is undoubtedly time-consuming and labor-intensive. 
Although some studies aim to ease model selection for non-expert developers, almost all of these approaches are either no longer maintained or focus narrowly on functional aspects while failing to account for other critical characteristics. 
Thus, there is an urgent need for a comprehensive approach that can flexibly account for all essential aspects in accordance with developers' requirements and automatically identify suitable PTMs for the developers with accuracy. 

In this study, we propose~\tool{}, an approach designed to accurately and efficiently identify potential PTMs that align with the diverse requirements of developers. 
In general, it is common and intuitive for developers to express their varied desired attributes of target models across distinct well-defined fields. 
By structuring developers intent in this way,~\tool{} is able to match the specified criteria against a curated set of pretrained models and accurately pinpoint the most qualified ones. 

%% file: Sections/approach.tex
\section{Preparation of Candidate PTMs}
Since~\tool{} aims to support the accurate and efficient selection of pretrained models for downstream developers, it is essential to prepare a representative and reliable set of candidate models. 
This collection provides a solid basis for conducting meaningful model selection and for rigorously evaluating the approach's effectiveness and generalizability across diverse tasks. 
In this section, we detail the collection and preprocessing steps undertaken to construct this candidate model dataset. 

\begin{figure}[!htbp]
  \centering
  \includegraphics[width=0.98\linewidth]{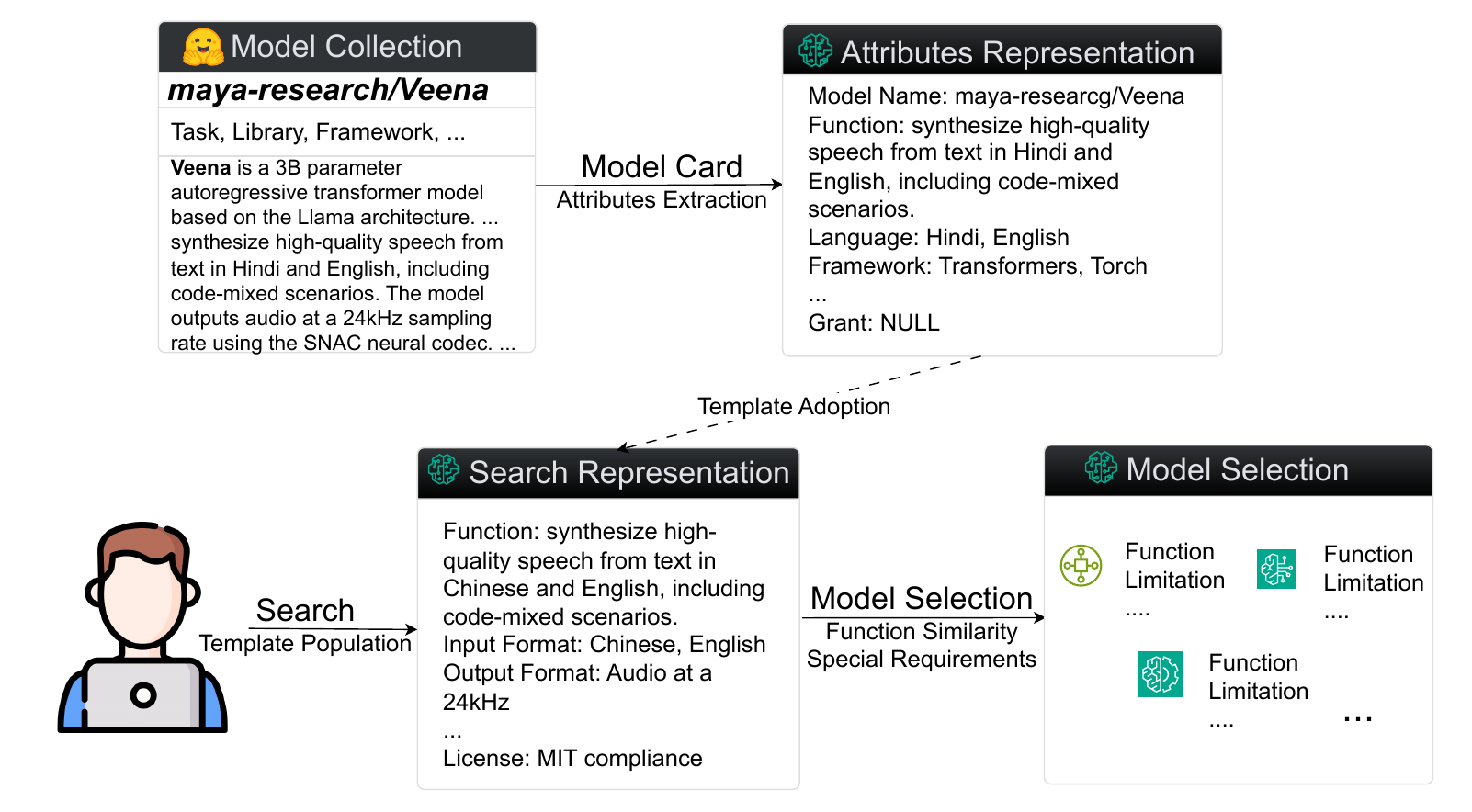}
  \caption{The workflow of~\tool{}.}
  \label{fig:workflow}
\end{figure}

The existing literature~\cite{jiang2022empirical,jiang2023ptmtorrent,jiang2024peatmoss} has examined several popular model hubs, such as Hugging Face~\cite{hfsite:2025}, Model Zoo~\cite{modelzoo:2025}, and ONNX Model Zoo~\cite{onnxmodel:2025}, which serve as hosting platforms for publicly available pretrained models (PTMs). 
In this study, Hugging Face is selected as the 
model hosting platform, as it offers the largest, most diverse, and best-documented collection of pretrained models compared to other hubs. 
Notably, our approach can be easily extended to support PTMs on other sharing platforms with minor modifications, enabling~\tool{} to integrate models from alternative sources. 

The component of ``Model Collection'' in Figure~\ref{fig:workflow} demonstrates an example of the publication of a pretrained model on Hugging Face, highlighting key elements such as the model registry name, relevant tags, and the model card. 
When model developers publish a pretrained model on Hugging Face with a registry name, they typically specify a set of descriptive tags, such as task (\ie{}, the model's designed purpose) and framework (\ie{}, the implementation platform). 
More importantly, the model publisher is also expected to provide a model card (a detailed specification of the model), which appears below the tags section. 
To streamline the creation of model cards, Hugging Face provides a standardized model card template~\cite{cardtemplate:2025} that can document key aspects, such as model function, training data, evaluation metrics, and interaction examples. 

\begin{figure}[!ht]
  \begin{minipage}[t]{0.46\linewidth}
    \centering
    \includegraphics[width=0.80\linewidth]{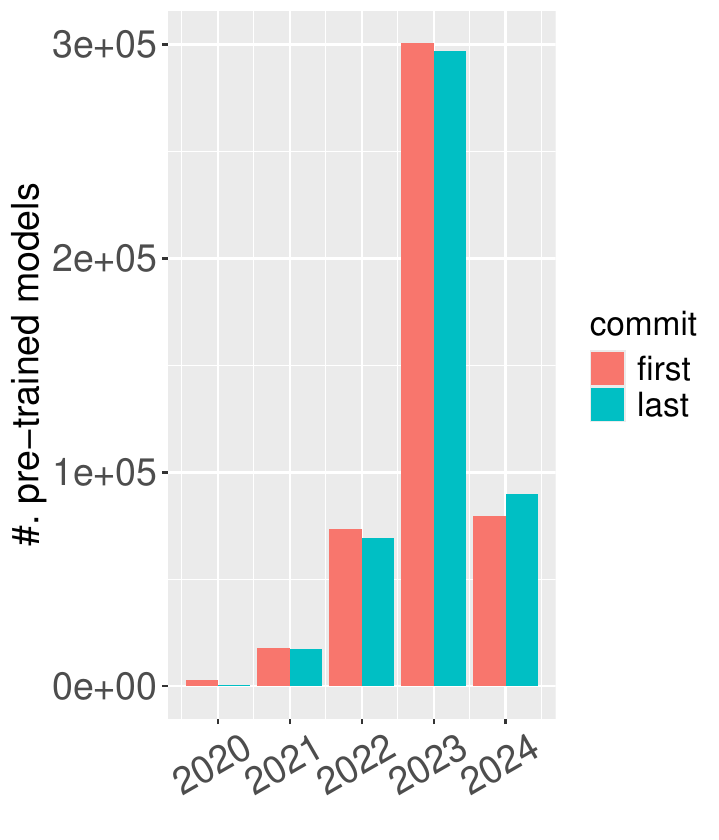}
    \caption{Number of pretrained models with respect to the year of first and last commit.} 
    \label{fig:modelcommit}
  \end{minipage}%
  \hspace{3mm}
  \begin{minipage}[t]{0.46\linewidth}
    \centering
    \includegraphics[width=0.80\linewidth]{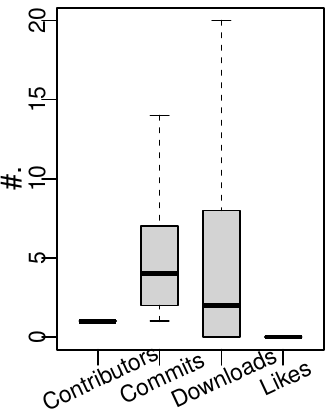}
    \caption{Number of contributors, commits, downloads, and likes of Hugging Face Models.}
    \label{fig:numberbasics}
  \end{minipage}
\end{figure}

The number of models hosted on Hugging Face is more than 544,000 as of the start date of our research. 
To efficiently scrape models hosted on Hugging Face, we resort to the popular Python web scraping framework, Scrapy~\cite{scrapy:2025}, which is an open-source and fast high-level web scraping and crawling framework licensed under BSD~\cite{bsd:2025}. 
Using the framework, we retrieved all available model registry entries and scraped all meaningful attributes for each model, including monthly downloads, user likes, and all other associated tags. 
In addition, we downloaded the associated model card by utilizing the~\texttt{huggingface\_hub}~\cite{hfdownload:2025} library for each of the collected model entries. 
In essence, the model card is a Markdown file (\texttt{README.md}) maintained in Git, and typically includes YAML front matter to describe metadata such as language, license, and task. 
Given the name of the model registry and the file extension (\eg{},~\texttt{.md}), the library reliably downloads all~\texttt{.md} files associated with the model, including the target model card. 

In this study, we successfully scraped a total of 543,949 models from Hugging Face\footnote{The number of hosted models is subject to frequent change and continues to grow rapidly.}. 
Figure~\ref{fig:modelcommit} presents the distribution of the models based on the year of their first and last commit. 
It clearly shows that the majority of pretrained models on Hugging Face were committed in 2023, which corresponds to the beginning of this study in early 2024. 
In contrast, only a limited number of models were committed in 2020 and 2021, corresponding to the early emergence of pretrained models. 

However, a substantial portion of these models are not considered valid because their model cards are either entirely missing or contain no content. 
A dedicated and well-documented model card for each pretrained model is essential for both application developers and AI practitioners to comprehensively understand its capabilities and limitations. 
The absence of valid model cards for models on Hugging Face suggests that these model authors may not prioritize usability and transparency for external users, but rather use the platform primarily as a hosting service. 
As a result, we obtained a total of 253,561 models, each accompanied by a valid model card.  
Among these, 87,022 had~\textit{zero} downloads in the past month, which may be because they are relatively unknown or fail to attract user interests.  
Figure~\ref{fig:numberbasics} presents a box plot of pretrained models with valid model cards. 
It illustrates the distribution of the number of contributors, commits, downloads in the past month, and likes received from external users. 
The median values for the number of contributors, commits, downloads in the past month, and likes are 1, 4, 2, and 0, respectively. 
The low median values for contributors and commits suggest that most models are maintained by an individual contributor and are infrequently updated. 
Similarly, the low median values for downloads and likes indicate that a substantial portion of pretrained models on Hugging Face receive limited attention from developers and researchers. 
This highlights the need for further investigation into the underlying causes and calls for greater efforts to improve model visibility and engagement. 



\section{\tool{}}
In this section, we present our approach,~\tool{}, which identifies suitable pretrained models for downstream application developers by leveraging structured well-defined fields organized according to a predefined template.  
Figure~\ref{fig:workflow} shows the overall workflow of~\tool{} for selecting pretrained models. 
With the predefined template,~\tool{} converts the candidate model specifications into well-structured fields and represents developers' requirements for target models in the same format. 
It then identifies suitable pretrained models by evaluating functional similarity and how well candidate models satisfy developer-specified special requirements. 
We now present the implementation details of~\tool{}, which provide the foundation for model selection. 


\begin{table}[!ht]
    \centering
    \caption{A list of summarized PTM metadata}
    \begin{threeparttable}
    \resizebox{0.98\linewidth}{!}{
    \begin{tabular}{c|l|l}
    \hline
       Source  &  \multicolumn{2}{c}{Metadata} \\
    \hline
     \multicolumn{1}{c|}{\multirow{24}{*}{Literature}} &  Framework  &  Underlying implementation framework  \\
    \cline{2-3}
                        & Copyright        &  Model copyright \\
    \cline{2-3}
                        & Evaluation       &  Evaluation dataset and related metrics\\
    \cline{2-3}
                        & Hardware         &  Training/inference hardware requirements \\
    \cline{2-3}
                        & Carbon emitted   &  Carbon emissions of the model\\
    \cline{2-3}
                        & Language         &  Languages supported by the model \\
    \cline{2-3}
                        & Software         &  Training/inference software requirements  \\
    \cline{2-3}
                        & Biases           &  The potential biases of the model \\
    \cline{2-3}
                        & Limitation       &  The limitations of the model \\
    \cline{2-3}
                        & Hyper-parameters &  The hyperparameter settings of the model \\ 
    \cline{2-3}
                        & Fine-tuning      &  Whether the model is fine-tuned \\
    \cline{2-3}
                        & Base model       &  Upstream base model name (if fine-tuned) \\
    \cline{2-3}
                        & Input format     &  The input of the model \\
    \cline{2-3}
                        & Grant            &  The grant or sponsor of the model \\
    \cline{2-3}
                        & Demo             &  Usage snippet or specification \\
    \cline{2-3}
                        & Report           &  Technical report/paper on the model \\
    \cline{2-3}
                        & Dataset          &  The model training or fine-tuning dataset \\
    \cline{2-3}
                        & Domain           &  The domain the model belongs to \\
    \cline{2-3}
                        & Inference cost   &  The model inference expenses \\
    \cline{2-3}
                        & Output format    &  The model output format \\
    \cline{2-3}
                        & Github repo      &  The source code the shared model \\
    \cline{2-3}
                        & Training cost    &  The model training expenses \\
    \cline{2-3}
                        & Model size       &  Local storage requirement \\
    \cline{2-3}
                        & Parameter size   &  The parameter size of the PTM \\
    \cline{2-3}
                        & Function         &  Brief summary of the model's function \\
    \cline{2-3}
                        & Others           &  Other training/inference requirements\\
    \hline
        Publicity Tags  & \multicolumn{2}{c}{\makecell{Model name, Dataset, Likes, Downloads\tnote{1}, \\ Library\tnote{2}, License, Contributors\tnote{3}, Commits\tnote{3}, Task}}\\
    \hline
    \end{tabular}
    }
    \begin{tablenotes}
    \small
    \item [1] The number of downloads last month
    \item [2] The implementation libraries of the model
    \item [3] The number of model contributors and commits
    \end{tablenotes}
    \end{threeparttable}
    \label{tab:modelmetadata}
\end{table}

\subsection{Metadata Attributes Representation for Candidate Models}


Although Hugging Face provides model card templates to encourage publishers to organize essential information into well-defined fields, in practice, many publishers still provide this information in unstructured free-text paragraphs. 
Therefore, gaining a comprehensive understanding of these models remains a significant challenge because key attributes are often buried in lengthy natural language descriptions. 
While such descriptions are interpretable by human readers, such as model researchers and downstream practitioners with extensive AI knowledge, they are not well-suited to developers with limited AI expertise nor to efficient batch processing by computers. 
Identifying suitable models still requires developers to manually sift through numerous model cards, which is both time-consuming and inefficient. 
In this study, we attempt to automatically extract essential and useful attributes from these unstructured natural language descriptions. 
This transformation streamlines the model selection process, enabling faster and more accurate identification of suitable models while reducing the manual burden (\ie{}, manual identification of key attributes from unstructured descriptions) for researchers and developers. 


Inspired by the existing literature~\cite{jiang2024peatmoss,schelter2017automatically,tsay2020aimmx,li2022metadata,tsay2022extracting}, we compiled the metadata attributes proposed for PTM descriptions and adapted extraction scripts provided in these studies to process the downloaded model cards. 
Table~\ref{tab:modelmetadata} presents the complete set of metadata fields examined in this study. 
The metadata attributes drawn from the literature refer to those proposed in related studies, and we provide brief explanations for them\footnote{Please refer to our artifact for a detailed explanation for each metadata.~\url{https://zenodo.org/records/15958951}}. 
In contrast, the metadata fields stemming from publicity tags are directly scraped from the model hosting platform and are generally considered self-explanatory.  
It is also worth noting that certain metadata fields may originate from both the~\texttt{Literature} and the~\texttt{Publicity Tags}. 
This is due to potential differences between attributes extracted from model cards and fields scraped directly from the model hosting publicity tags. 
With regard to duplicated metadata fields, we consolidated them into a single representation to avoid redundancy. 
In total, we have 33 distinct metadata attributes that constitute our predefined~\textit{template} for describing both models and developer-specified requirements. 

\begin{figure*}[h]
  \centering
  \includegraphics[width=0.98\linewidth]{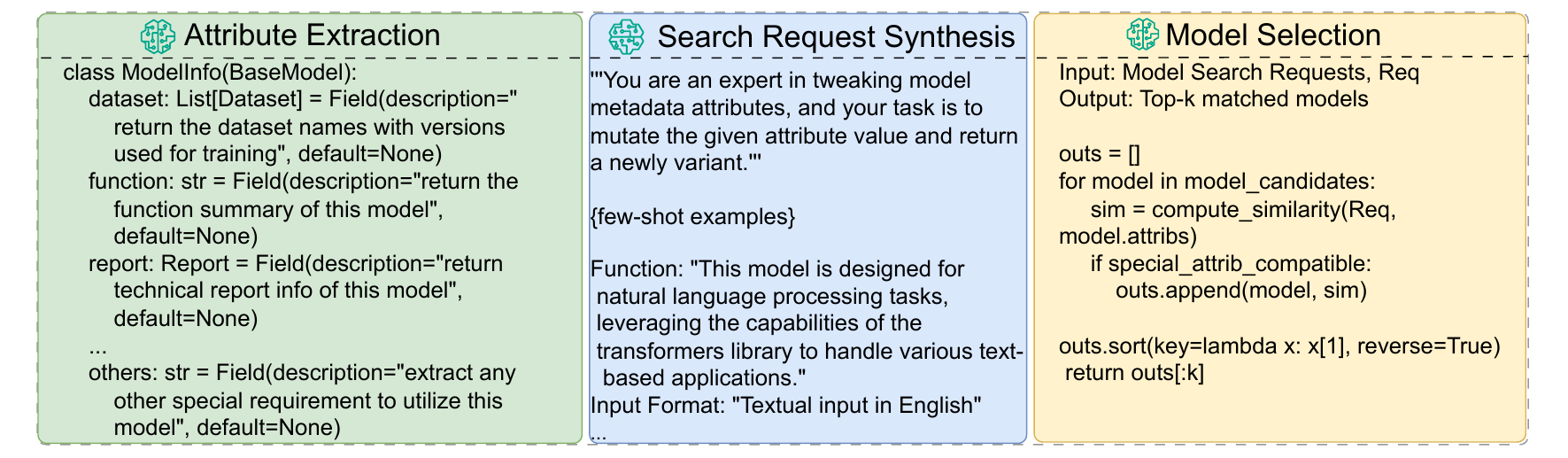}
  \caption{Prompt design in the process of dataset preparation and model selection.}
  \label{fig:modelprompt}
\end{figure*}

\begin{figure*}
\begin{minipage}{.96\textwidth}
\begin{lstlisting}[
    caption=The example of metadata attributes of a curated pretrained model,
    label={lst:promptgen},
    language=C++,
    backgroundcolor=\color{backcolour},   
    commentstyle=\color{codegreen},
    keywordstyle=\color{magenta},
    numberstyle=\tiny\color{codegray},
    stringstyle=\color{codepurple},
    basicstyle=\ttfamily\footnotesize,
    breakatwhitespace=false,         
    breaklines=true,                 
    keepspaces=true,                 
    numbers=left,       
    numbersep=5pt,                  
    showspaces=false,                
    showstringspaces=false,
    showtabs=false,                  
    tabsize=1,
    ]
{"function": "This model is designed for question answering tasks specifically tailored for the Arabic language, leveraging the capabilities of the AraELECTRA architecture.", 
"input_format": "Preprocessed text in Arabic for context and question", 
"output_format": "Extracted answer with confidence score and position indices", 
"task": "question-answering",
"license": "Apache-2.0",
"framework": "Transformer"}
\end{lstlisting}
\end{minipage}\hfill
\begin{minipage}{.96\textwidth}
\begin{lstlisting}[
    caption=Model search structured request (template instance) mutation example,
    label={lst:promptmu},
    language=C++,
    backgroundcolor=\color{backcolour},   
    commentstyle=\color{codegreen},
    keywordstyle=\color{magenta},
    numberstyle=\tiny\color{codegray},
    stringstyle=\color{codepurple},
    basicstyle=\ttfamily\footnotesize,
    breakatwhitespace=false,         
    breaklines=true,                 
    keepspaces=true,                 
    numbers=left,       
    numbersep=5pt,                  
    showspaces=false,                
    showstringspaces=false,
    showtabs=false,                  
    tabsize=1,
    ]
{"Input": {
  "function": "Model for Mongolian speech recognition using the Common Voice dataset.",
  "input_format": "Audio file at 16kHz",
  "output_format": "Transcribed text",
  "task": "automatic-speech-recognition"}}
{"Output": {
  "function": "Mongolian text-to-speech synthesis", 
  "input_format": "Text in Mongolian",
  "output_format": "Audio file at 16kHz",
  "task": "text-to-speech" }}
\end{lstlisting}  
\end{minipage}
\end{figure*}

To effectively and efficiently extract metadata attributes from scraped model cards, we referred to an existing artifact~\cite{mdextract:2025,jiang2024peatmoss}, which achieves an attribute extraction accuracy of up to 94.39\%.
We adapted this artifact for the specific objectives of our research, as our work requires extracting a broader set of metadata attributes than the original implementation. 
Furthermore, we leveraged recent advances in large language models by incorporating a more reliable and capable LLM (\ie{}, GPT-4~\cite{gpt4:2025} in this study). 
Consequently, we were able to effectively extract the metadata attributes, as shown in the~\texttt{Metadata Extraction} component of Figure~\ref{fig:modelprompt}. 
Listing~\ref{lst:promptgen} illustrates extracted metadata attributes for a pretrained model, including function, input\_format, output\_format, and task. 
In practice, the extraction process is implemented using the Python library Instructor~\cite{pyinstructor:2025}, a widely adopted tool for structuring the outputs of underlying LLMs, such as Mixtral~\cite{mixtral:2025} and GPT-4~\cite{gpt4:2025}. 

\subsection{Model Search Requests Synthesis}
\label{subsec:modelsearch}
In practice, it is common for developers to express the desired characteristics of target models through well-defined structured fields, such as the template described above. 
Thus, in the model selection stage of~\tool{}, the desired model attributes specified by developers can be compared with the corresponding attributes of candidate PTMs to determine whether each candidate model meets the intended requirements. 
This shift from traditional keyword-based searching to detailed attribute-level comparison marks a significant advancement in model selection, ensuring that all relevant factors are systematically considered in the model search decision-making process. 
Unfortunately, there is no readily available dataset consisting of developer-specified attribute requirements for target models (also referred to as developer-specified model search requests) that are structured according to the predefined template described above. 
To address this, in this research, we curated our own model search request dataset. 

To construct this dataset, we populated the predefined template with attribute values synthesized using large language models. 
The generation of these values follows mutation strategies based on prior work~\cite{patil2023gorilla,zhuo2024bigcodebench,jia2010analysis,papadakis2019mutation} and is guided by the metadata values extracted from the curated pretrained models. 
By leveraging the extracted attribute values and few-shot prompting, we designed effective prompts to generate diverse and semantically meaningful attribute values, as shown in the~\texttt{Search Request Synthesis} component of Figure~\ref{fig:modelprompt}.  
For example, we can relax or tighten the license compatibility requirements, modify the input format (\eg{}, changing the input language or modality), or adjust the core functionality (\eg{}, shifting from text-to-image generation to image-to-text generation). 
Specifically, in terms of attributes with a limited value set, such as license, input\_format, and output\_format, we substitute their original values with semantically valid alternatives. 
Regarding attributes, such as function and limitation, we employ prompt-based LLMs with few-shot examples and mask-and-infill techniques to modify key semantic elements while preserving the fluency and coherence of the original descriptions. 
Listing~\ref{lst:promptmu} presents an example of model search request mutation. 
Harnessing the generative capabilities of large language models, we can effectively mutate given model-populated instances, thereby producing new, diverse, and semantically valid model search requests. 
Therefore, this approach not only broadens the scope of template-based model search requests but also improves the flexibility and adaptability of the search process, enabling it to accommodate more sophisticated model search demands.  

\begin{algorithm}
\caption{Model Selection}
\label{alg:cap}
\begin{algorithmic}[1]
\Input Model search requests~\textbf{Req} and candidate models~\textbf{C}
\Output Top-K matched models~\textbf{M}
\Attributes [license, copyright, hardware, software, training\_cost, inference\_cost, limitation, biases, size] 
\State $req.attribs \gets ~$\textbf{Req} \Comment{Splitting the search request (search template instance) attributes into special and other attributes}
\For{\textit{c} in ~\textbf{C}}
    \State $c.sim \gets ~$embedding\_similarity(req.attribs\_others, \textit{c}.attribs\_others) 
    \For{attrib in req.attribs\_special}
        \If{attrib not compatible with \textit{c}.attribs\_special}
            \State $Break $
        \EndIf
    \EndFor
    \If{$c.sim >~$sim\_threshold}
        \State $\textbf{M}.append([\textit{c}, \textit{c}.sim])$
    \EndIf
\EndFor
\State $\textbf{M}.sort(key = sim)$
\end{algorithmic}
\end{algorithm}

\subsection{Model Selection}
Equipped with diverse model search requests and a valid candidate PTM dataset, we are now able to perform model selection. 
The ``Model Selection'' component in Figure~\ref{fig:modelprompt} illustrates the general steps involved in the selection process, while Algorithm~\ref{alg:cap} further describes the detailed procedure to identify the target models. 
To select appropriate models, the search request attributes are first categorized into two distinct groups---Special and Trivial (\ie{}, line 1 in Algorithm~\ref{alg:cap}). 
The attributes are categorized into two groups based on whether they can be straightforwardly determined to align with the attribute values of candidate models. 
This distinction is critical because it determines the strategy for identifying satisfied attributes, which in turn leads to selecting the appropriate models. 

Special attributes are those that require more nuanced handling and cannot be directly compared with the developer-specified attribute values, while trivial ones are all the remaining attributes, which do not need special consideration and can be used for comparison straightforwardly.  
For example, the trivial attribute~\texttt{function}, which briefly summarizes the model's primary purpose, can be directly used to perform an accurate text match between the function attribute in the model search request and that in the candidate model-populated instances. 
Specifically, for trivial attributes, the embedding similarity based on BM25~\cite{robertson2009probabilistic}, though not limited to it, is computed between the attribute values extracted from the candidate models and those specified in the model search requests. 
This semantic similarity comparison helps to achieve a more efficient matching process. 

When dealing with special attributes, such as license and bias, attribute-specific strategies are necessary. 
For instance, suppose that a developer requests a model that must be compatible with the MIT license. 
Candidate models with licenses such as Apache 2.0 and BSD both can meet the license requirement, as these licenses are generally permissive. 
In contrast, models under the GPT license conflict with the requirement. 
A direct and literal match based on the special attributes may result in many other potentially suitable models being overlooked. 
To accurately assess these special attributes, we designed a set of customized prompts, each tailored to address the unique nature of the corresponding attribute.  
These prompts ensure that these special attributes are properly considered. 
As a result, the reliability of the model selection process is significantly enhanced, reducing the risk of selecting models that meet the main functional requirements but do not satisfy the special constraints specified in the developer's request. 

To identify appropriate models, we rank candidate models based on their calculated similarity values and filter out those that do not satisfy the developers' special requirements. 
This two-step approach, involving specific prompts for special attributes and embedding similarity measurements for trivial attributes, ensures that the model selection process is both precise and efficient.

%% file: Sections/evaluation.tex
\section{Evaluation}
The operation of~\tool{} hinges on the availability of well-defined model search requests and curated candidate pretrained models. 
More specifically, the final selection performance depends on the quality of these inputs, including accurate metadata extraction from candidate pretrained models and the provision of high-quality search requests. 
Thus, to provide a comprehensive evaluation of~\tool{}, we propose the following four research questions in this section: the first three examine metadata extraction and search request synthesis, while the fourth evaluates the effectiveness of the final model selection. 
Furthermore, we detail the experimental dataset and the underlying configuration of the advanced large language model (\ie{}, GPT-4). 

\subsection{Research Questions}
Metadata extraction from pretrained models and model search request synthesis: 
\begin{itemize}
    \item RQ1: How reliable is GPT-4 in extracting metadata from model cards?
    \item RQ2: What are the key characteristics of the metadata associated with candidate models?
    \item RQ3: How reliable is the synthesis of mutated model search requests? 
\end{itemize}
Model selection: 
\begin{itemize}
    \item RQ4: How accurate is~\tool{} in selecting pretrained models? 
\end{itemize}

\subsection{Experimental Dataset}
In this study, we collected a total of 543,949 pretrained models from Hugging Face, among which 253,561 are considered valid since they are accompanied by non-empty model cards. 
However, a notable proportion of these valid models (\ie{}, 87,022) had the number of downloads last month~\textit{zero}, suggesting limited visibility and the potential omission of essential information. 
To ensure that well-maintained and widely adopted models are included in our evaluation, we applied a stricter filtering criterion based on model popularity. 
We raised the minimum threshold for downloads last month to 2,000 to curate the experimental dataset and obtained a final set of 5,069 pretrained models. 
\vspace{-3pt}
\subsection{Evaluation Setup}
Addressing these research questions requires both extracting metadata from model card descriptions and subsequently synthesizing structured model search requests using the extracted information. 
Thanks to rapid advances in large language models~\cite{mixtral:2025,gpt4:2025}, both metadata extraction and search request synthesis can be performed effectively and efficiently through well-designed prompts. 
In this study, we selected GPT-4~\cite{gpt4:2025} as our experimental LLM because of its state-of-the-art capabilities and widespread adoption in the software engineering research community. 
Specifically, we carefully designed the prompts shown in Figure~\ref{fig:modelprompt} and provided three input-output examples to elicit the model to generate more concise and accurate results. 
We used the default 4k context window and retained all default GPT-4 settings except for the temperature and TopP parameters, which were manually set to 0.0 and 0.95, respectively.
This configuration minimizes randomness in the output while preserving sufficient creativity to support effective extraction and generation of analysis results. 

\subsection{Evaluation Results}
\subsubsection{RQ1: Reliability of metadata extraction} \hfill\\
To effectively extract the metadata fields listed in Table~\ref{tab:modelmetadata} and better align with the specific objectives of our study, we adapted the existing model card extraction scripts\footnote{\url{https://github.com/PurdueDualityLab/PeaTMOSS-Artifact/tree/main/LLM-Pipeline/Accurate_Pipeline}}, which have shown high efficiency and achieved an extraction accuracy of up to 94.39\%, thus providing a solid foundation for our investigation. 
Specifically, we extended the original functionality to support a broader set of metadata attributes and enhanced the extraction pipeline by integrating a more recent version of GPT-4, which offers greater capabilities and reliability compared to the original implementation. 
This research question, thus, evaluates the effectiveness and reliability of metadata extraction. 

However, automatically assessing these aspects is challenging due to the absence of ground truth, thereby necessitating manual verification. 
Moreover, it is impractical for us to manually inspect the metadata extraction from all 5,069 experimental PTMs, as this would be excessively time-consuming. 
Therefore, we randomly sampled and manually evaluated 100 models from the curated PTM collection, referencing the evaluation approach in~\cite{jiang2024peatmoss}. 

Specifically, two independent authors conducted this evaluation. 
Each independently reviewed the original model cards of the pretrained models to assess whether the extracted values accurately reflected the information presented in the source documents. 
Consequently, discrepancies or inconsistencies in the evaluation of certain metadata values could naturally arise. 
In such cases, meetings were convened in which both authors engaged to resolve any discrepancies until consensus was reached. 
By conducting this rigorous review and reconciliation process, we obtained a subjective evaluation of our metadata extraction procedure, which serves as the basis for the subsequent steps toward model selection.  

Finally, two independent authors achieved high inter-rater reliability, as measured by Cohen's unweighted kappa coefficient with the value of 0.81~\cite{cohenkappa}, which means almost perfect agreement between two raters.  
As a result of this process, the accuracy of metadata extraction based on the manually inspected sample of 100 model cards reached 95\%.  
To complement and automate this manual inspection, we enhanced the validation process by crafting specialized prompts with few-shot examples to assess the validity of the metadata extraction. 
These prompts were designed to simulate the human validation process, in which the correctness of extracted metadata is assessed against the corresponding model card. 
Specifically, the few-shot examples embedded in these prompts guided the language model to more accurately assess the validity of each extracted attribute. 
Ultimately, this prompt-based validation approach achieved an accuracy of 95\%, matching the manual inspection results and demonstrating its effectiveness as a scalable alternative. 

\begin{tcolorbox}[before skip=0.4cm, after skip=0.6cm, title=\textbf{RQ1 Findings}, left=2pt, right=2pt,top=2pt,bottom=2pt]
The advancement of large language models, when paired with carefully designed extraction prompts, enables highly accurate metadata extraction, reaching an impressive accuracy of 95\%. 
\end{tcolorbox}

\subsubsection{RQ2: Characteristics of metadata} \hfill\\
\begin{figure}
    \centering
    \includegraphics[width=0.9\linewidth]{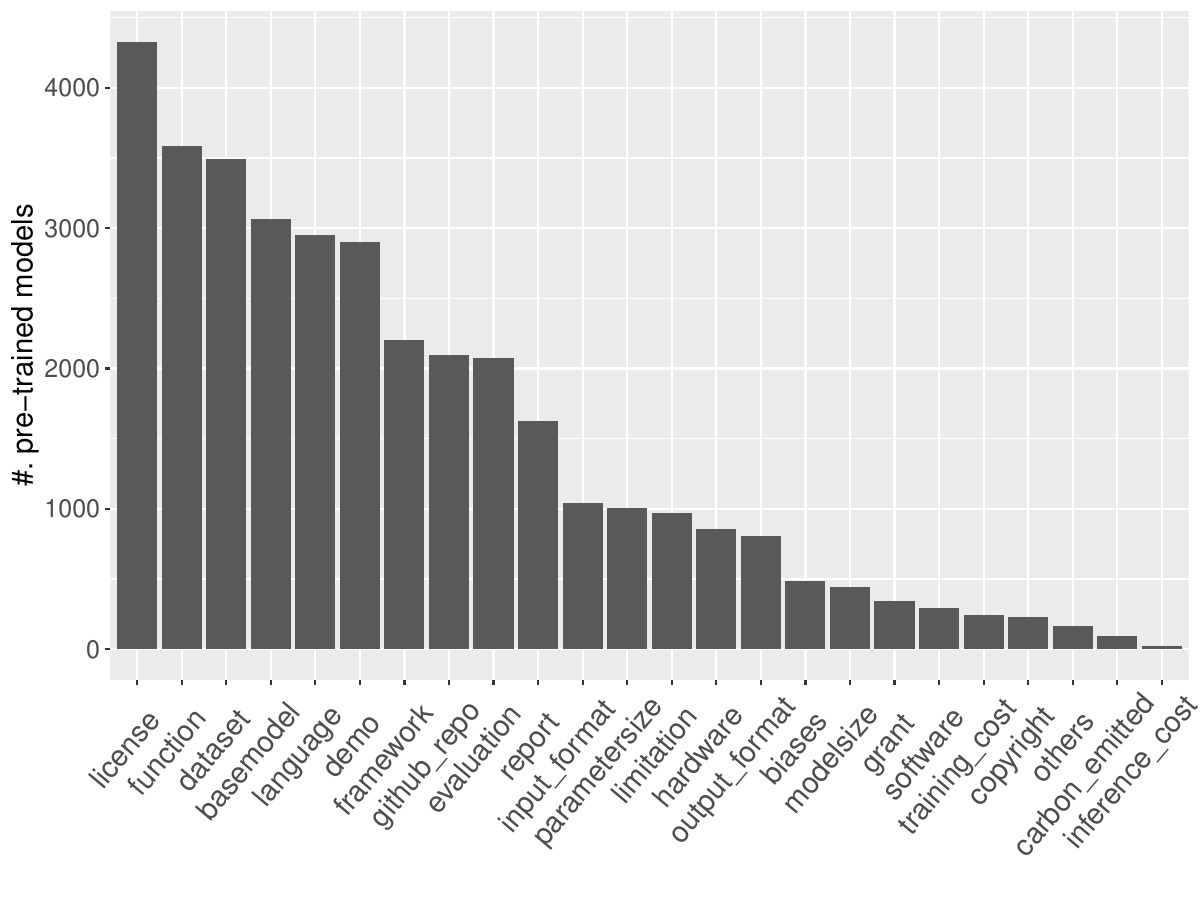}
    \caption{The number of models regarding each extracted metadata.}
    \label{fig:mdcount}
\end{figure}
A detailed analysis of the extracted metadata provides critical insights into its richness, structure, and representativeness, informing its utility in supporting effective model search and selection. 
Moreover, this perspective reveals which characteristics of pretrained models are emphasized by developers and highlights patterns that can guide future model design by identifying underrepresented or overlooked attributes. 
Consequently, this research question examines the extracted metadata to gain detailed insights into its characteristics.  

\begin{figure*}[!ht]
  \begin{minipage}[t]{0.94\linewidth}
    \centering
    \includegraphics[width=0.94\linewidth]{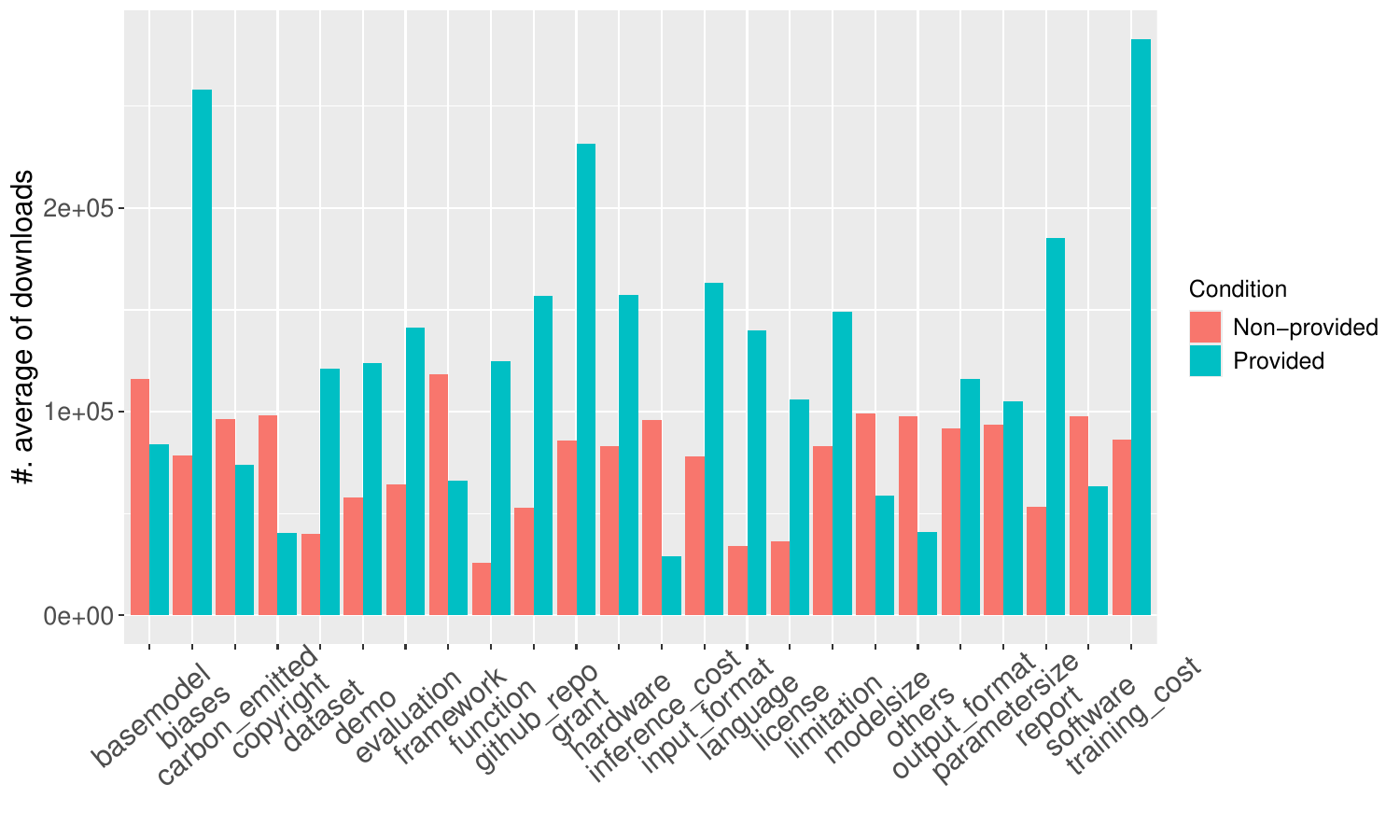}
    \caption{The average number of downloads regarding different metadata.} 
    \label{fig:mdavg}
  \end{minipage}%
  \vspace{3mm}
  \begin{minipage}[t]{0.94\linewidth}
    \centering
    \includegraphics[width=0.94\linewidth]{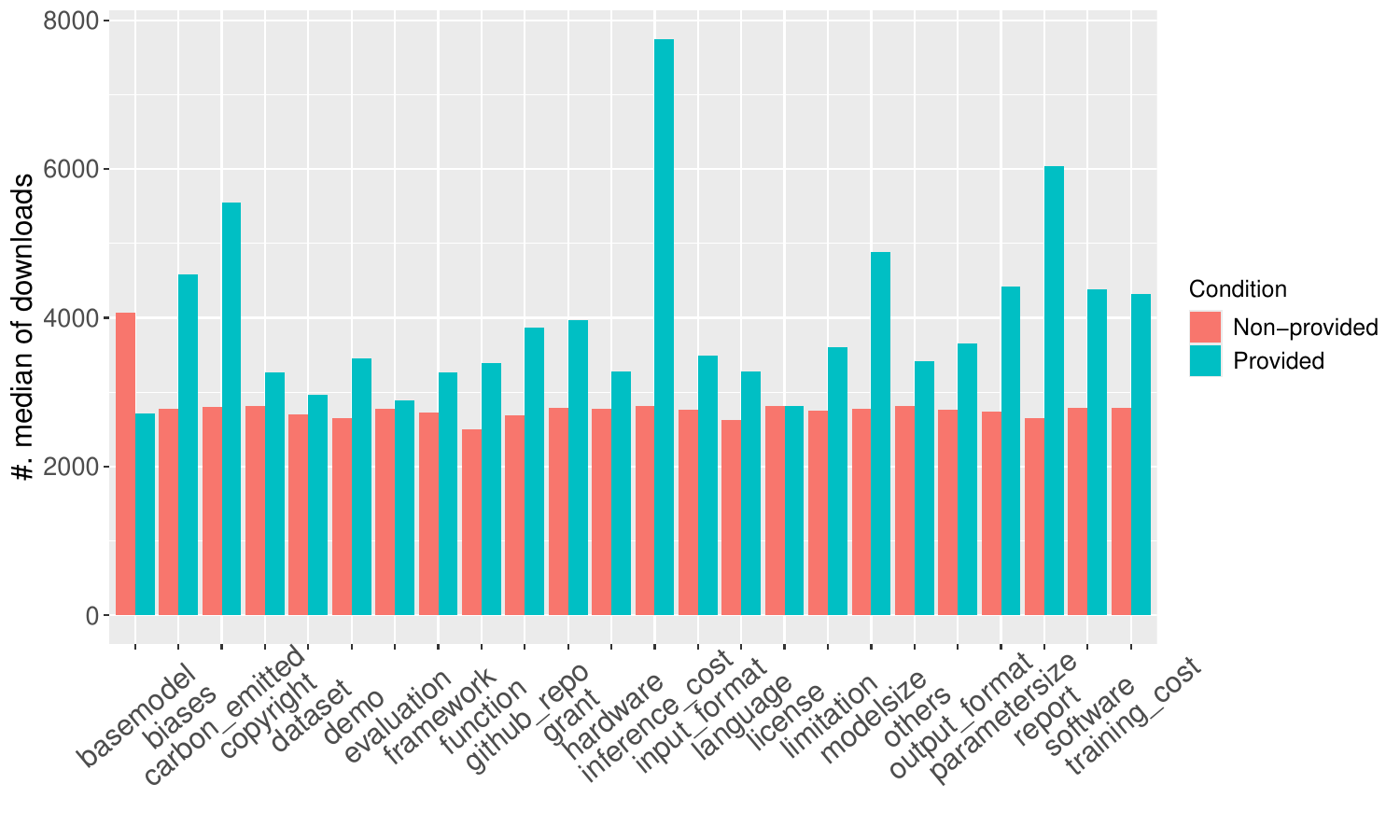}
    \caption{The median number of downloads regarding different metadata.}
    \label{fig:mdmedian}
  \end{minipage}
\end{figure*}

Fig.\ref{fig:mdcount} presents the number of models associated with each summarized metadata attribute. 
The three most frequently provided metadata fields among the selected PTMs are~\textit{license},~\textit{function}, and~\textit{dataset}. 
Among these, the \textit{function} is extracted from the model's associated model card using GPT-4 with carefully crafted prompts, while the others can be directly specified by model developers through the Hugging Face metadata UI or by editing the associated YAML file\cite{modelmd:2025} and thus are straightforwardly scraped from the model hosting site. 
In contrast, most of the remaining metadata fields are not explicitly required by the hosting platform, and thus can only be extracted from the free-form content of the model card.  
This suggests that, in the absence of mandatory fields, authors are generally less inclined to voluntarily provide detailed metadata. 

Fig.~\ref{fig:mdavg} and Fig.~\ref{fig:mdmedian} present the average and median downloads for each extracted metadata field, depending on whether the metadata field is introduced in the model card. 
In general, PTMs associated with certain metadata tend to exhibit higher downloads, a trend also observed in prior work~\cite{jones2024we}. 
This correlation suggests that the presence of certain metadata improves the visibility, perceived quality, or usability of PTMs, thereby contributing to their increased adoptions. 
However, certain metadata fields, including basemodel, framework, software, copyright, modelsize, inference cost, and carbon emitted, exhibit an opposite trend, where PTMs associated with these attributes tend to have fewer downloads. 
One possible explanation is that users may perceive these metadata fields as less critical or less directly relevant to their application needs. 
For instance, metadata such as the model's base model (\ie{}, the upstream model used for pretraining or fine-tuning) may be regarded as implementation-specific and of limited relevance to functional performance.
Additionally, the infrequent provision of these specific metadata may contribute to their lower perceived importance and, consequently, to fewer downloads. 

In summary, while the majority of the presented metadata generally correlates with higher downloads, there are exceptions where a small portion of the attributes seems to deter users. 
This observation indicates that not all metadata are equally valued and their perceived utility may vary depending on context and user needs.
However, to accommodate diverse application requirements and enhance model visibility, we advocate for PTM authors to provide as much detailed metadata as possible. 

\begin{tcolorbox}[before skip=0.4cm, after skip=0.6cm, title=\textbf{RQ2 Findings}, left=2pt, right=2pt,top=2pt,bottom=2pt]
PTM authors often provide only the information explicitly required by the template, which can result in omitting other valuable metadata. 
In general, model cards with more comprehensive metadata generally attract higher download counts, indicating greater visibility and value. 
To enhance discoverability and adoption, PTM authors are encouraged to include as much relevant details as possible. 
\end{tcolorbox}

\subsubsection{RQ3: Reliability of model search requests synthesis} \hfill\\
The performance of~\tool{} is shaped not only by the curated pool of pretrained models but also by the availability and relevance of model search requests. 
Since no such dataset of model search requests exists, we constructed one to support our study. 
During this process, three additional model search requests were synthesized for each curated PTM, derived from its original metadata representation and in accordance with the procedure described in Section~\ref{subsec:modelsearch}, resulting in a total of 15,207 $(5,069 \times 3)$ model search requests. 

To assess the reliability of the synthesized model search requests, we randomly selected 100 pretrained models from the experimental dataset and synthesized three additional distinct model search requests for each, yielding a total of 300 $(100 \times 3)$ synthesized requests. 
Subsequently, a multi-step validation process was employed. 
Specifically, GPT-4 was first engaged using carefully crafted prompts to assess the validity of each request and to produce comprehensive justifications for its validation determinations. 

Beyond the automated validation, two independent authors then reviewed GPT-4's final determinations and explanations to cross-validate the model's outcomes. 
For instance, the function specification ``Model for facial recognition and attribute analysis'' and the license requirement ``license: Apache-2.0'' in the search request were correctly classified as valid. 
The detailed explanation stated: ``The model search request is valid. The user is looking for a model that performs facial recognition and attribute analysis, which are compatible tasks. The requirement for the model to be under the Apache-2.0 license is also clear and non-conflicting. Therefore, there are no conflicting requirements in this request''. 

In contrast, there were also cases of incorrect validation. 
For example, the function requirement ``Model for code translation between programming languages'' and the license specification ``license CC-BY-NC-4.0'' were incorrectly classified as invalid. 
The explanation stated that ``The CC BY-NC 4.0 license is a Creative Commons license that restricts use to non-commercial purposes only. This means that any model found under this license cannot be used for commercial purposes, which is a significant restriction. Depending on the intended use, especially if the requester plans to use the model for commercial projects, this could be a conflicting requirement. Therefore, the request is considered invalid due to the restrictive nature of the CC BY-NC 4.0 license in the context of potential commercial use.''. 
This explanation, however, is incorrect, as we do not consider downstream application constraints at this stage and instead focus solely on whether there are conflicts within the search request itself.  
After an in-depth analysis of the sampled validation results, the independent authors reached a consensus that the accuracy of the model search request generation and mutation process was as high as 84.67\%. 
This finding underscores the effectiveness of using large language models such as GPT-4 to enhance the model selection process by ensuring that valid and coherent search requests are synthesized. 

\begin{tcolorbox}[before skip=0.4cm, after skip=0.6cm, title=\textbf{RQ3 Findings}, left=2pt, right=2pt,top=2pt,bottom=2pt]
With carefully crafted prompts, the advanced LLM (GPT-4) employed in this study achieved a high accuracy of 84.67\% in synthesizing valid model search requests. 
This result underscores the effectiveness of our prompt-driven synthesis strategy in producing structurally sound and conflict-free search requests.
\end{tcolorbox}
\vspace{-5pt}

\subsubsection{RQ4: Accuracy of model selection} \hfill\\
With a large-scale dataset of pretrained models and a high-quality of model search requests in place, we are able to address the core research question of whether~\tool{} is accurate and reliable in selecting pretrained models for developers. 
However, a significant challenge remains: the absence of ground truth data makes it difficult to accurately measure the performance and reliability of our model selection approach,~\tool{}. 

To address this gap, we incorporate manual evaluation alongside our automated model selection process. 
Specifically, we engage human experts to assess the outputs of~\tool{} and determine whether they adequately meet the intended requirements of developers. 
By relying on human judgment, we aim to validate the accuracy and reliability of the system, thereby compensating for the absence of predefined ground truth in this domain.

In the evaluation process, human experts are provided with both the model search queries and the candidate PTMs, which are ranked in descending order based on the computed similarity scores, as described in Algorithm~\ref{alg:cap}. 
For each search query, the reviewers examined the top 10 candidate PTMs and determined whether each candidate fully met the requirements, partially satisfied the requirements, or did not meet the requirements at all. 
Candidate PTMs that fully meet the requirements are defined as those that satisfy every aspect specified in the model search request. 
In contrast, partially satisfied models are those that fulfill only some of the specified tasks or requirements but fall short in other areas. 
For example, the function specification ``Model for object detection and segmentation in real-time video streams'' and the license requirement ``license: Apache-2.0'' in the model search request may identify a pretrained model with the function description ``Object detection model trained to detect forklifts and persons''.
However, this model only partially fulfills the request, as it addresses object detection for specific objects, but does not perform segmentation. 

\begin{figure}
    \centering
    \includegraphics[width=0.98\linewidth]{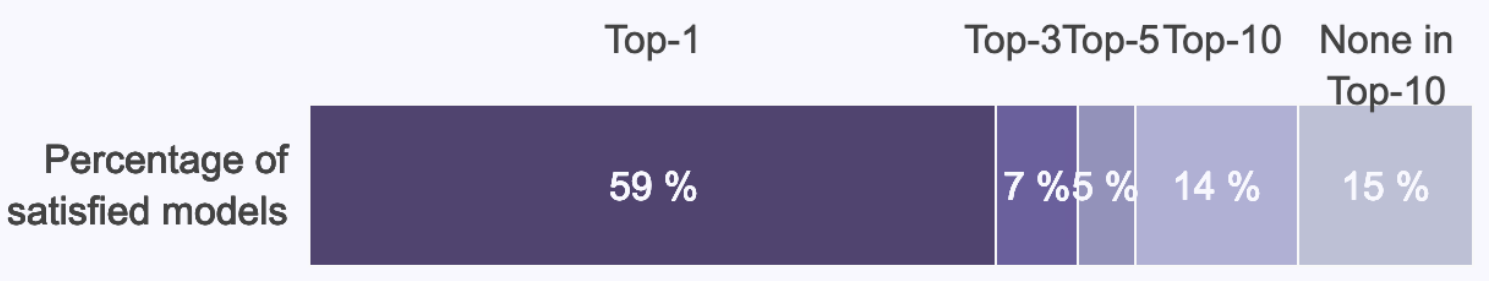}
    \caption{Evaluation results for pretrained model selection.}
    \label{fig:eval}
\end{figure}

To make the evaluation process manageable, two independent authors manually verified a random sample of 100 valid synthesized model search requests drawn from the subset of valid requests within the 300 previously generated. 
Furthermore, we excluded the 100 PTMs previously used to generate the 300 model search requests to prevent overlap. 
With the valid model search requests and candidate models prepared,~\tool{} performed the model selection to identify suitable pretrained models for each request. 
Figure~\ref{fig:eval} summarizes the results of the manual inspection.  
Out of the 100 sampled requests, 85\% successfully identified appropriate PTMs within the top 10 candidates, demonstrating the effectiveness of our approach in facilitating suitable model selection. 
Specifically, among these 85 requests, 59 found fully satisfying PTMs in the top-ranked candidate, 66 identified fully satisfying PTMs within the top three candidates, and 71 did so within the top five candidates. 
Additionally, 11 of the 85 requests identified models that only partially fulfilled the search requirements within the top 10 candidates. 
For the remaining 15 sampled requests, no appropriate PTMs were found within the top 10 candidates. 
For instance, the function requirement ``Experimental model fine-tuned on Hercules v3 for text generation tasks with input as text and output as textual response'', combined with the license specification, requires that the candidate model should be fine-tuned on the Hercules v3 dataset. 
However, no appropriate model using Hercules v3 was found among the top 10 candidates. 
The identified PTMs with higher similarity values only provide a generic function description such as ``Text generation model fine-tuned for various tasks'', which does not meet the specified requirement regarding the fine-tuning dataset. 

\begin{tcolorbox}[before skip=0.4cm, after skip=0.6cm, title=\textbf{RQ4 Findings}, left=2pt, right=2pt,top=2pt,bottom=2pt]
Manual evaluation on the sampled experimental dataset shows that~\tool{} is a powerful and effective approach for selecting suitable pretrained models, achieving a high model selection accuracy of 85\%. 
\end{tcolorbox}



%% file: Sections/discussion.tex
\section{Discussion}
\subsection{Implications}
\label{sec:implication}

\textbf{Model developers and maintainers.}
Developers and maintainers of pretrained models should make as much effort as they can to provide extensive and detailed information about their published models. 
Our research has demonstrated a clear positive correlation between the amount of available information and the frequency of downloads. 
The more comprehensive the documentation, including usage examples, clear description of model input and output, model limitations, the more likely users are engaged with the model, resulting in a higher number of downloads. 
Additionally, to facilitate efficient model selection for downstream application developers and enable effective batch processing by computers, model developers could present detailed information in a key-value pair format. 

\textbf{Downstream application developers.}  
Relying on superficial metrics, such as downloads and likes, application developers may overlook pretrained models that better meet their project requirements. 
Therefore, they should not only consider these digital metrics but also pay close attention to the detailed descriptions provided by model developers. 
In addition, application developers could learn from this research and express their specific functional requirements and constraints for target models from the perspective of model developers, thereby achieving a more precise match with suitable pretrained models.  
\vspace{-4pt}
\subsection{Threats to Validity}
\label{sec:threats}
\textbf{External validity} The primary external validity lies in the dataset construction. 
In this study, all curated models hosted on Hugging Face follow the same standardized publication template. 
PTMs hosted on a different hosting site may not have the same standard, which can make metadata extraction ineffective, even invalid. 
However, the metadata summarized in this study is still general and representative. 
Thanks to the advance of Large Language Models, only small changes in the metadata extraction prompt could mitigate this threat. 
In addition, we evaluated the accuracy and reliability of our approach using a small sample of synthesized requests and pretrained models, which may not fully represent the entire experimental dataset. 
However, the experimental requests and models were randomly selected, which helps mitigate this threat to validity to some extent.

\textbf{Internal validity} The main concerns of internal validity are the outputs of the underlying LLM (\ie{}, GPT-4 in this research) and the participation of humans to assess the proposed approach. 
In terms of LLM, the inherent hallucination and inconsistent output from multiple executions cannot be avoided, which may also be preferred by LLMs (\ie{}, creative knowledge generation). 
In this work, to mitigate this threat, the authors of the paper iteratively discussed and manually fine-tuned these prompts to ensure valid and accurate results. 
Regarding the involvement of human evaluators in assessing accuracy and reliability, some degree of bias in manual inspection is inevitable. 
To mitigate this threat, two authors independently cross-validated the results and discussed discrepancies to reach a final consensus.